\documentclass[12pt]{iopart}
\usepackage{graphicx}

\begin{document}

\title{Inseparability inequalities for higher-order moments for bipartite systems}

\author{G. S. Agarwal$^1$\footnote{On leave of absence from Physical Research
Laboratory, Navrangpura, Ahmedabad, India} and Asoka Biswas$^2$}
\address{$^1$Department of
Physics, Oklahoma state University, Stillwater, OK - 74078, USA}
\address{$^2$Physical Research Laboratory, Navrangpura, Ahmedabad - 380
009, India} \eads{\mailto{agirish@okstate.edu},
\mailto{asoka@prl.ernet.in}}
\date{\today}

\begin{abstract}
There are several examples of bipartite entangled states of
continuous variables for which the existing criteria for
entanglement using the inequalities involving the second order
moments are insufficient. We derive new inequalities involving
higher order correlation, for testing entanglement in non-Gaussian
states. In this context we study an example of a non-Gaussian
state, which is a bipartite entangled state of the form
$\psi(x_{\rm a},x_{\rm b})\propto (\alpha x_{\rm a}+\beta x_{\rm
b})e^{-(x_{\rm a}^2+x_{\rm b}^2)/2}$. Our results open up an
avenue to search for new inequalities to test entanglement in
non-Gaussian states.
\end{abstract}


\maketitle

The detection and characterization of entanglement in the state of
a composite is an important issue in quantum information science.
Peres has addressed this issue \cite{peres} for the first time to
show that inseparability of a bipartite composite system can be
understood in terms of negative eigenvalues of partial transpose
of its density operator. There are several other criteria for
inseparability in terms of correlation entropy and linear entropy
\cite{chuang,huang_gsa} and in terms of positivity of
Glauber-Sudarshan $P$-function \cite{kim}. However all these
measures cannot be put to experimental tests. To detect
entanglement of any composite system experimentally, one needs to
have certain criteria in terms of expectation values of some
observables.

Using Peres's criterion of separability, Simon \cite{simon} has
derived certain separability inequalities, violation of which is
sufficient to detect entanglement in bipartite systems. These
inequalities involve variances of relative position and total
momentum coordinates of the two subsystems and thus can be
verified experimentally \cite{howell}. Duan et al. \cite{duan}
have also derived equivalent inequalities independently using the
positivity of the quadratic forms. It is further proved that for
Gaussian states (states with Gaussian wave functions in coordinate
space), violation of these inequalities provides a necessary and
sufficient criterion for entanglement. A different form of
criterion for entanglement involving second order moments has been
derived by Mancini \emph{et al.} \cite{mancini}. These
inequalities have been tested for entangled states produced by
optical parametric oscillators and other systems where the output
state can be approximated by Gaussian states
\cite{bowen,silberhorn,josse,polzik}.

In context of quantum information and communication, non-Gaussian
states are equally important as Gaussian states. Several entangled
non-Gaussian states have been studied in literature
\cite{gsa_puri,kim,grangier}. A way to produce non-Gaussian state
is via state reduction method \cite{gsa_qo,yamamoto,bellini}. Thus
characterization of entanglement in non-Gaussian state remains an
open question. This motivates us to derive new inequalities, when
the existing inequalities based on second order correlation fail
to test entanglement in these states. Thus these inequalities are
expected to involve higher order correlation between position and
momentum coordinates. In this paper we consider a bipartite
entangled state of bosonic system, which, in turn, is a
non-Gaussian state in coordinate space. We consider an entangled
state for which the existing inseparability inequalities cannot
provide any information about the inseparability of the state. We
derive new inseparability inequalities to test its entanglement.

We start by deriving the inequalities involving the second order
moments. Consider the set of operators
\begin{equation}
U=\frac{1}{\sqrt{2}}(x_{\rm a}+x_{\rm
b})~;~V=\frac{1}{\sqrt{2}}(p_{\rm a}+p_{\rm b}) \qquad [U,V]=i\;.
\end{equation}
Then we would have the uncertainty relation
\begin{equation}
\Delta U\Delta V\ge \frac{1}{2}\;.
\end{equation}
We now use Peres-Horodecki criteria of separability in terms of
the partial transpose. Under partial transpose, $x_{\rm
b}\rightarrow x_{\rm b}$, $p_{\rm b}\rightarrow -p_{\rm b}$. Hence
the condition that the partial transpose of a density matrix is
also a genuine density matrix, would imply that
\begin{equation}
\Delta\left(\frac{x_{\rm a}+x_{\rm
b}}{\sqrt{2}}\right)\Delta\left(\frac{p_{\rm a}-p_{\rm
b}}{\sqrt{2}}\right)\ge \frac{1}{2}\;. \label{mancini}
\end{equation}
This inequality was first derived by Mancini \emph{et al.}
\cite{mancini} by using a very different method. Thus if a
bipartite system is separable then (\ref{mancini}) should be
obeyed. Violation of \eref{mancini} gives a sufficient condition
for entanglement. The inequality of Duan et al. follow from
(\ref{mancini}) by using the relations
\begin{equation}
\eqalign{M^2=M_-^2+4M_{\rm x}\;,\cr M=[\langle(\Delta
u)^2\rangle+\langle(\Delta v)^2\rangle]\;,\cr M_-=[\langle(\Delta
u)^2\rangle-\langle(\Delta v)^2\rangle]\;, \cr M_{\rm
x}=\langle(\Delta u)^2\rangle\langle(\Delta v)^2\rangle\;,\cr}
\label{duan_mancini}
\end{equation}
where
\begin{equation}
u=x_{\rm a}+x_{\rm b}\;,\quad\mathrm{and}\quad v=p_{\rm a}-p_{\rm
b}\;.
\end{equation}
From \Eref{duan_mancini} it is clear that if the inequality
$M_{\rm x}\geq 1$ (which is the separability criterion of Mancini
{\it et al.\/}) holds, then the criterion $M\geq 2$ which is the
separability criterion of Duan {\it et al.\/}, is automatically
satisfied for all values of $M_-$. But if $M_{\rm x}<1$, then
nothing can be said about the exact value of $M$. It can be
greater than or less than 2 depending on the values of $M_{\rm x}$
and $M_-$. The above analysis implies that the separability
criterion given by Mancini {\it et al.\/} and that given by Duan
{\it et al.\/} are interrelated with each other. Furthermore,
$M_{\rm x}<1$ is stronger than the criterion $M<2$ for
inseparability. This follows from $M_{\rm x}\le M/2$. We also note
that Duan \emph{et al.} derived a more general separability
inequality
\begin{equation}
\left|m^2-\frac{1}{m^2}\right|\leq M<m^2+\frac{1}{m^2},
\label{duan1a}
\end{equation}
where
\begin{equation}
u=|m|x_{\rm a}+\frac{1}{m}x_{\rm b}\;,\qquad v=|m|p_{\rm
a}-\frac{1}{m}p_{\rm b}\;. \label{uv}
\end{equation}

\noindent For a bipartite Gaussian states, the inequalities for
second order correlations are also sufficient. Equivalent
necessary and sufficient condition for separability of Gaussian
states have been derived by Englert and Wodkiewicz \cite{englert}
using density operator formalism. They have shown that the
positivity of the partial transposition and P-representability of
the separable Gaussian states are closely related.

In this paper we focus on the following bipartite continuous
variable Bell state formed from ground and excited states of the
harmonic oscillators:
\begin{equation}
\psi(x_{\rm a},x_{\rm b})=\sqrt{\frac{2}{\pi}}(\alpha x_{\rm
a}+\beta x_{\rm b})e^{-(x_{\rm a}^2+x_{\rm b}^2)/2}\;,\qquad
|\alpha|^2+|\beta|^2=1\;, \label{state}
\end{equation}
which is the state of a composite system of bosonic particles. It
clearly represents a non-Gaussian state in coordinate space. The
nonclassical properties of such states were studied in
\cite{gsa_puri}. A recent experimental proposal discusses how to
generate non-Gaussian states by subtracting a photon from each
mode of a two-mode squeezed vacuum state \cite{grangier}

The Peres-Horodecki criterion \cite{peres} is known to be
necessary and sufficient for inseparability for bipartite systems
in $(2\times 2)$ and $(2\times 3)$ dimensions, but to be only
sufficient for any higher dimensions. This criterion states that
if the partial transpose of a bipartite density matrix has at
least one negative eigenvalue, then the state must be inseparable.
Next, we apply this criterion to test the inseparability of the
state (\ref{state}). The density matrix of this state is given by
\begin{equation}
\eqalign{\rho &= |\psi\rangle\langle\psi|\nonumber \cr
&=|\alpha|^2|1,0\rangle\langle 1,0|+|\beta|^2|0,1\rangle\langle
0,1|+(\alpha^*\beta|0,1\rangle\langle 1,0|+\mathrm{H.c.})\;,}
\end{equation}
where
\begin{equation}
|1,0\rangle \equiv \sqrt{\frac{2}{\pi}}x_{\rm a}e^{-(x_{\rm
a}^2+x_{\rm b}^2)/2}\;.
\end{equation}
Taking partial transpose of the second subsystem, we obtain the
following density matrix,
\begin{equation}
\eqalign{\rho^{\mathrm{PT}}=|\alpha|^2|1,0\rangle\langle
1,0|+|\beta|^2|0,1\rangle\langle 0,1|
+(\alpha^*\beta|0,0\rangle\langle 1,1|+\mathrm{H.c.})\;.}
\end{equation}
The four eigenvalues of the above density matrix can be calculated
as $|\alpha|^2$, $|\beta|^2$, $\pm|\alpha||\beta|$. Clearly the
negative eigenvalue of $\rho^{\mathrm{PT}}$ confirms the
inseparability of the state (\ref{state}) under consideration.

Now we examine the validity of the existing inseparability
inequalities (\ref{duan1a}) and (\ref{mancini}) for the entangled
state (\ref{state}). For the conjugate variables $u$ and $v$
defined by \Eref{uv}, we find that
\begin{equation}
\langle (\Delta u)^2\rangle+\langle (\Delta
v)^2\rangle=|m|^2+\frac{1}{m^2}+2\left(|\alpha|^2|m|^2+\frac{1}{m^2}|\beta|^2\right)\;,
\end{equation}
which is clearly greater than $|m|^2+1/m^2$. Thus though the state
(\ref{state}) is entangled, the criterion (\ref{duan1a}) cannot
exploit this fact. In other words, violation of the criterion
(\ref{duan1a}), as shown above, would conclude that the state
under consideration is separable, which definitely is not the
case. We further find that for $m=1$,
\begin{equation}
\langle (\Delta u)^2\rangle\langle (\Delta
v)^2\rangle=4-(\alpha\beta^*+\alpha^*\beta)^2=4-4[\mathrm{Re}(\alpha\beta^*)]^2\;,
\end{equation}
which has the minimum value equal to 3, which implies that
$\langle (\Delta u)^2\rangle\langle (\Delta v)^2\rangle$ is always
greater than unity for $m=1$. According to the inequality
(\ref{mancini}), this refers to separability in the state which is
again not the case. From the above discussion we conclude that the
existing inseparability criteria based on second order
correlations do not provide correct information about the
inseparability of a standard bipartite entangled state which in
turn is non-Gaussian. This warrants search for new inequalities
involving higher order correlations, to test inseparability of
such states.

Needless to say that since there is an infinity of these
higher-order correlations, one could construct a very large number
of such inequalities involving higher order correlations. In what
follows, we consider the next logical correlations. Our analysis
below is reminiscence of what has been done in context of
nonclassical light \cite{hillery,nonclass}. We could consider the
following set of operators:
\begin{equation}
S_{\rm x}=\frac{a^\dag b+ab^\dag}{2};~S_{\rm y}=\frac{a^\dag
b-ab^\dag}{2i},~S_{\rm z}=\frac{a^\dag a-b^\dag b}{2}\;.
\end{equation}
The operators $S_i$ obey the algebra of angular momentum operators
and hence the uncertainty relation $\Delta S_{\rm x}\Delta S_{\rm
y}\ge \frac{1}{2}|\langle S_{\rm z}\rangle|$ would give, for
example,
\begin{equation}
\Delta\left[\frac{a^\dag
b+ab^\dag}{2}\right]\Delta\left[\frac{a^\dag
b-ab^\dag}{2i}\right]\ge \frac{1}{2}\left|\left\langle\frac{a^\dag
a-b^\dag b}{2}\right\rangle\right|\;. \label{pt1}
\end{equation}
It is known that under partial transpose, a separable density
matrix remains as a valid density operator. Using this property of
partial transpose, we expect that for a separable state, the
following inequality is also valid:
\begin{equation}
\fl\eqalign{[\langle a^\dag abb^\dag\rangle+\langle aa^\dag b^\dag
b\rangle+\langle {a^\dag}^2 {b^\dag}^2\rangle+\langle a^2
b^2\rangle -\langle a^\dag b^\dag +ab\rangle^2]\cr \times [\langle
a^\dag abb^\dag\rangle+\langle aa^\dag b^\dag b\rangle-\langle
{a^\dag}^2 {b^\dag}^2\rangle-\langle a^2 b^2\rangle +\langle
a^\dag b^\dag -ab\rangle^2] \geq \left|\langle a^\dag a-b^\dag
b\rangle\right|^2} \label{pt}
\end{equation}
which has been obtained from \Eref{pt1} under the partial
transpose $b\leftrightarrow b^\dag$. A violation of (\ref{pt})
would imply that the state is entangled. However, for the state
(\ref{state}), $|(|\alpha|^2-|\beta|^2)|\le 1$. Thus the
inequality (\ref{pt}) is not violated and hence does not lead to
any new information regarding the inseparability of the state.
Next we consider the following operators satisfying SU(1,1)
algebra
\begin{equation}
K_{\rm x}=\frac{a^\dag b^\dag+ab}{2};~K_{\rm y}=\frac{a^\dag
b^\dag-ab}{2i},~K_{\rm z}=\frac{a^\dag a+b^\dag b+1}{2}\;.
\end{equation}
Such operators previously have been used in consideration of
higher order squeezing \cite{hillery}. The uncertainty inequality
would give
\begin{equation}
\Delta\left[\frac{a^\dag
b^\dag+ab}{2}\right]\Delta\left[\frac{a^\dag
b^\dag-ab}{2i}\right]\ge \frac{1}{2}\left|\left\langle\frac{a^\dag
a+b^\dag b+1}{2}\right\rangle\right|\;.
\end{equation}
Using the partial transpose as above, we get a new inequality for
separability
\begin{equation}
\fl\eqalign{[\langle a^\dag ab^\dag b\rangle+\langle aa^\dag
bb^\dag\rangle+\langle {a^\dag}^2 {b}^2\rangle+\langle a^2
{b^\dag}^2\rangle -\langle a^\dag b +ab^\dag\rangle^2]\cr \times
[\langle a^\dag ab^\dag b\rangle+\langle aa^\dag bb^\dag
\rangle-\langle {a^\dag}^2 {b}^2\rangle-\langle a^2
{b^\dag}^2\rangle +\langle a^\dag b -ab^\dag\rangle^2]\geq
\left|\langle a^\dag a+bb^\dag \rangle\right|^2\;.} \label{eq1}
\end{equation}
For the state (\ref{state}), the above relation leads to the
following result:
\begin{equation}
\left|\alpha^*\beta\right|^2-2[\mathrm{Re}(\alpha^*\beta)]^2[\mathrm{Im}(\alpha^*\beta)]^2\leq
0
\end{equation}
which is always violated for all values of $\alpha$ and $\beta$.
Thus the state under consideration is inseparable according to
this inequality (\ref{eq1}) which is in conformity with
Peres-Horodecki criterion. This inequality, which is based on
higher order correlation, is thus  successful to test
inseparability in the non-Gaussian states like (\ref{state}),
while the existing inequalities based on second order correlation
fail to do so. This result opens up an avenue to search for
general inseparability inequalities for non-Gaussian states.

Note that the inequality (\ref{eq1}) can also be expressed in
terms of position and momentum variables of the two subsystems as
\begin{eqnarray}
\left[\Delta^2(x_{\rm a}x_{\rm b})+\Delta^2(p_{\rm a}p_{\rm
b})+\langle x_{\rm a}p_{\rm a}p_{\rm b}x_{\rm b}\rangle+\langle
p_{\rm a}x_{\rm a}x_{\rm b}p_{\rm b}\rangle-2\langle
x_{\rm a}x_{\rm b}\rangle\langle p_{\rm a}p_{\rm b}\rangle\right]\nonumber\\
\times \left[\Delta^2(x_{\rm a}p_{\rm b})+\Delta^2(p_{\rm a}x_{\rm
b})-\langle x_{\rm a}p_{\rm a}x_{\rm b}p_{\rm b}\rangle-\langle
p_{\rm a}x_{\rm a}p_{\rm b}x_{\rm b}\rangle+2\langle
x_{\rm a}p_{\rm b}\rangle\langle p_{\rm a}x_{\rm b}\rangle\right]\nonumber\\
\hspace{1cm}\geq\frac{1}{4}\left|\langle x_{\rm
a}^2\rangle+\langle p_{\rm a}^2\rangle+\langle x_{\rm b}^2\rangle
+\langle p_{\rm b}^2\rangle\right|^2 \label{eq2}
\end{eqnarray}

In order to detect entanglement using \Eref{eq1} we need to do a
variety of homodyne measurements
\cite{raymer,leonhardt,kumar,gsa_chatur}. Such measurements would
yield the distribution of quadratures.


In conclusions, we have shown in context of a bosonic non-Gaussian
state of the Bell form, that the existing inseparability
inequalities based on second order correlations are not enough to
test the entanglement. We have derived a new set of separability
inequalities using Peres-Horodecki criterion of separability. This
new inequality involves higher order correlation of quadrature
variables and can be tested experimentally as discussed above.
Violation of this inequality detects entanglement in the
non-Gaussian state under consideration. The failure of the
existing criteria in terms of the second order moments is perhaps
a reflection of the fact that the state (\ref{state}) in no limit
goes over to a Gaussian state.

\end{document}